\documentclass[aps,pre,onecolumn,showkeys,groupedaddress]{revtex4-2}
\usepackage{color, amsmath, amssymb, bm, graphicx, tikz}

\newcommand{\pfrac}[2]{\frac{\partial #1}{\partial #2}}
\newcommand{\e}{\mathrm{e}}
\newcommand{\ii}{\mathrm{i}}
\newcommand{\figpath}{}

\begin{document}

\title{Stability of similarity solutions of viscous thread pinch-off}

\author{Michael C. Dallaston}
\affiliation{School of Mathematical Sciences, Queensland University of Technology, Brisbane, QLD, 4001, Australia}

\author{Chengxi Zhao\protect\footnote{Current Address: Department of Modern Mechanics, University of Science and Technology of China, Hefei 230026, China.}}
\affiliation{School of Engineering, University of Warwick, Coventry CV4 7AL, United Kingdom}

\author{James E. Sprittles}
\affiliation{Mathematics Institute, University of Warwick, Coventry CV4 7AL, United Kingdom}

\author{Jens Eggers}
\affiliation{School of Mathematics, University of Bristol, Bristol BS8 1UG, United Kingdom}

\date{\today}

\begin{abstract}
In this paper we compute the linear stability of similarity solutions of the breakup of viscous liquid threads, in which the viscosity and inertia of the liquid are in balance with the surface tension.  The stability of the similarity solution is determined using numerical continuation to find the dominant eigenvalues. Stability of the first two solutions (those with largest minimum  radius) is considered.  We find that the first similarity solution, which is the one seen in experiments and simulations, is linearly stable with a complex nontrivial eigenvalue, which could explain the phenomenon of break-up producing sequences of small satellite droplets of decreasing radius near a main pinch-off point.  The second solution is seen to be linearly unstable.  These linear stability results compare favorably to numerical simulations for the stable similarity solution, while a profile starting near the unstable similarity solution is shown to very rapidly leave the linear regime.

\end{abstract}

\maketitle

\section{Introduction}
\label{sec: intro}

Liquid threads or jets consist of cylindrical liquid matter, bounded by an interface with another fluid.  The interfacial energy (that is, surface tension) on such an interface  makes it unstable to perturbations that result in the thread breaking up; a typical everyday example is a column of water pouring from a kitchen tap forming into droplets some distance from the tap, but there are countless applications in science and industry across orders of magnitude of spatial scales, from the formation of galaxies, to pharmaceutical drug delivery, and inkjet printing \citep{Eggers2008}.  In many industrial applications, the breakup of the liquid thread is either undesirable, or the aim is to control the breakup to create a field of very fine droplets of a similar size (a monodisperse field) \citep{Barrero2007,Ganan2001}.  In the latter case, the dynamics of breakup are very important, as complex break-up phenomena, for instance the creation of satellite and sub-satellite droplets, can result in a wide distribution of droplet sizes.

A cylindrical thread or jet of liquid is unstable to interfacial perturbations due to the Plateau--Rayleigh instability, driven by surface tension \citep{Eggers1997,Eggers2008}.  Eventually non-linear effects take over and the liquid breaks up into droplets via pinch-off singularities, at which the thread radius locally goes to zero.  Both numerical simulations and experiments indicate that universal self-similarity plays a crucial role near pinch-off.  However, the dynamics near pinch-off can be complex, including iterated structures and break-up into droplets of multiple sizes \citep{Brenner1994,Latikka2020}.  The nature of the self-similarity depends on the relative importance of effects such as the inertia and the viscosity of the ambient fluid.  In a single simulation or experiment, multiple regimes may be observed \citep{Castrejon2015,Li2016,Lagarde2018}, with a complex picture of transitions across multiple regimes tangled up with seemingly oscillatory convergence towards similarity solutions.

If the outer fluid and inertia are neglected, so that the dominant effects are viscosity of the inner fluid and surface tension, then exact solutions exist for the countably infinite number of similarity solutions \citep{Eggers1997,Papageorgiou1995}.  The linear stability of these solutions may also be determined exactly; in \citet{Eggers2012} it was shown that of the infinite number of solutions, only one such solution (that with the largest minimum radius) is stable, while all other solutions are unstable.  Indeed, it is common more widely for partial differential equation (PDE) models that have a countable set of similarity solutions or other invariant solutions for only one to be linearly stable: see, for example \citet{Witelski1999} in the case of self-similar van der Waals rupture of thin films, \citet{Bernoff1998} in the case of pinch-off by surface diffusion, or \citet{Kessler1985,Bensimon1986,Tanveer1987} in the case of travelling Saffman--Taylor fingers in Hele--Shaw flow. 

If, however, inertia and viscosity are both in balance with surface tension, similarity solutions can only be computed numerically.  In addition, the computation is challenging because of a singular point in the equation, at which an extra condition is imposed by requiring that solutions be analytic.  In the past, shooting methods that utilize a power series expansion at the singular point have been used to successfully compute the first similarity solution \citep{Eggers1993,Eggers1995} that agrees with numerical simulation of the PDEs, as well as a sequence of secondary similarity solutions, that are not seen in the numerical simulations \citep{Brenner1996}.  \citet{Brenner1996} have conjectured that for viscous thread break-up with inertia, all similarity solutions are linearly stable, and a difference in sensitivity to finite-size perturbations is why the first similarity solution is the one that is generically seen in full numerical simulations.

\begin{figure}
\centering
\includegraphics{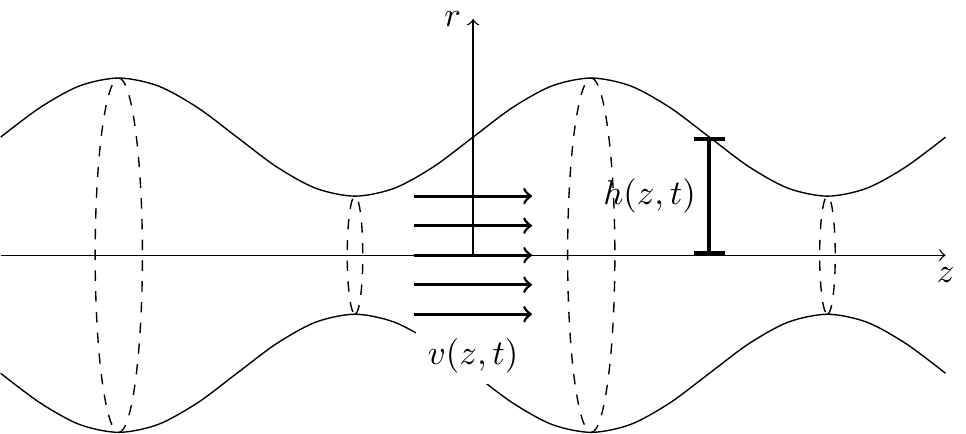}

\caption{Schematic of a thin thread of viscous fluid.  The system is modelled with two variables: the streamwise velocity $v(z,t)$ (uniform across the thread in the slender body approximation) and the thread radius $h(z,t)$, each functions of the streamwise coordinate $z$ and time $t$.}
\end{figure}

In this article we examine the linear stability of these similarity solutions.  As with the computation of the similarity profiles themselves, the linear stability is computationally challenging due to the presence of the singular point, and has never been attempted.  Previously, numerical continuation has been successfully used to compute similarity solutions of the thin film equation \citep{Tseluiko2013,Dallaston2017} and determine their stability \citep{Dallaston2018}.  In this article we use numerical continuation, starting from solutions to the equations on smaller domains generated by shooting, to construct the first two similarity solutions for viscous-inertial thread break-up, and find the most unstable (nontrivial) eigenvalue to determine their stability.  Our chief finding is that the base similarity solution is linearly stable, but the second is linearly unstable, counter to the prediction by \citet{Brenner1996}. In both cases the dominant eigenvalue is complex.  For the stable solution, the effect of the dominant eigenvalue matches quantitatively with what is seen in numerical computation of the PDE system, correctly predicting the frequency and decay rate of the perturbation around the base similarity solution.

We proceed by describing the formulation of the slender body Navier Stokes equations used to model the evolution of the radius and fluid velocity of the liquid thread in Section \ref{sec:formulation}, then in Section \ref{sec:similarity} describe the numerical continuation procedure used to find both the similarity solutions and the relevant dominant eigenvalues that determine their stability.  In Section \ref{sec:numerics} we show how the stability results match with carefully performed numerical simulations.  Conclusions are presented in Section \ref{sec:discussion}.

\section{Formulation and similarity solutions}
\label{sec:formulation}

After applying nondimensionalization and slender body theory to the Navier--Stokes equation inside the jet, with kinematic and surface tension conditions on the interface, the system is modelled using two partial differential equations describing the jet radius $h(z,t)$, and fluid velocity $v(z,t)$, as a function of position $z$ and time $t$:
\begin{subequations}
\label{eqsys:pdes}
\begin{align}
\pfrac{h}{t} + v\pfrac{h}{z} &= -\frac{h}{2}\pfrac{v}{z} \label{eq:pde1} \\
\pfrac{v}{t} + v\pfrac{v}{z} &= -\pfrac{}{z}\left(\frac{1}{h(1+h_z^2)^{1/2}} - \frac{h_{zz}}{(1+h_z^2)^{3/2}}\right) + \frac{1}{h^2}\pfrac{}{z}\left(h^2\pfrac{v}{z}\right) \label{eq:pde2}.
\end{align}
\end{subequations}
(see for example \citet{Eggers1993}).  Here we have excluded terms such as gravity that are negligible in the similarity analysis near pinch-off, but have retained higher order contributions from the surface tension, which we include in the numerical computation of solutions to (\ref{eqsys:pdes}) in Section \ref{sec:numerics}.  The system is parameter free as any implicit parameters are in the initial and boundary conditions, which do not play a role in universal similarity solutions near pinch-off.

The jet pinches off when $h \to 0$ at a point $z = z_0$ and finite time $t=t_0$.  We seek similarity solutions by choosing a reference frame that `zooms in' on such a point\citep{Giga1985}, by defining
\begin{equation}
v(z,t) = (t_0-t)^{-1/2}\psi(\xi, \tau), \quad h(z,t) = (t_0-t)\phi(\xi,\tau), \quad \xi = \frac{z-z_0}{(t_0-t)^{1/2}}, \quad \tau = -\log(t_0 - t).
\end{equation}
This substitution results in the system
\begin{subequations}
\label{eqsys:scaledpdes}
\begin{equation}
\phi_\tau + (\psi + \xi/2)\phi_\xi = (1-\psi_\xi/2)\phi \label{eq:scaledpde1}
\end{equation}
for (\ref{eq:pde1}) and
\begin{equation}
\psi_\tau + \psi/2 + \xi\psi_\xi/2 + \psi\psi_\xi = 
-\left[\frac{1}{\phi(1+\e^{-\tau}\phi_\xi^2)^{1/2}} - \frac{\e^{-\tau}\phi_{\xi\xi}}{(1+\e^{-\tau}\phi_\xi^2)^{3/2}}\right]_\xi + \frac{1}{\phi^2}\left[\phi^2\psi_\xi\right]_\xi. \nonumber
\end{equation}
for (\ref{eq:pde2}).  Note that the terms resulting from higher order contributions from the surface tension in this equation are order $t_0-t = \e^{-\tau}$.  These will prove to be negligible compared to the dynamics of interest in the limit $t\to t_0$, resulting in
\begin{equation}
\psi_\tau + \psi/2 + \xi\psi_\xi/2 + \psi\psi_\xi = \frac{\phi_\xi}{\phi^2} + 3\psi_{\xi\xi} + 
6\frac{\psi_\xi\phi_\xi}{\phi}.
\label{eq:scaledpde2}
\end{equation}
\end{subequations}
Steady states ($\phi_\tau = \psi_\tau = 0$) of (\ref{eqsys:scaledpdes}) satisfy the same system of ordinary differential equations that result from assuming $\phi$ and $\psi$ are functions of $\xi$ only, and thus are similarity solutions of the original system (\ref{eqsys:pdes}).  Including the dependence on the logarithmic time variable $\tau$ allows us to examine the stability of these steady states, to determine which is seen in practice, and the manner in which any such stable similarity solution is approached.

\section{Computation of similarity solutions and eigenvalues}
\label{sec:similarity}

To simplify the algebra slightly, we substitute $\omega = \log\phi$, so that the system (\ref{eqsys:scaledpdes}) is
\begin{subequations}
\label{eqsys:pde}
\begin{align}
\omega_\tau + (\psi + \xi/2)\omega_\xi &= (1-\psi_\xi/2) \\
\psi_\tau + \psi/2 + \xi\psi_\xi/2 + \psi\psi_\xi &= \e^{-\omega}\omega_\xi + 3\psi_{\xi\xi} + 6\psi_\xi\omega_\xi.
\end{align}
\end{subequations}
To investigate the linear stability, write $\omega(\xi,\tau) = \bar \omega(\xi) + \tilde\omega(\xi)\e^{\sigma\tau}$, and $\psi(\xi,\tau) = \bar \psi(\xi) + \tilde\psi(\xi)\e^{\sigma\tau}$, where $\sigma$ is the eigenvalue.  The base state, that is, the similarity solutions themeselves, satisfy
\begin{subequations}
\label{eqsys:base0}
\begin{align}
(\bar\psi + \xi/2)\bar \omega_\xi &= (1-\bar \psi'/2) \\
\bar\psi/2 + \xi\bar\psi'/2 + \bar\psi\bar\psi' &= \e^{-\bar\omega}\bar\omega' + 3\bar\psi'' + 6\bar\psi'\omega',
\end{align}
\end{subequations}
and the equations for the perturbations are
\begin{subequations}
\label{eqsys:corrn0}
\begin{align}
\sigma\tilde\omega + (\bar\psi+\xi/2)\tilde\omega' + \tilde\psi\bar\omega' &= -\tilde\psi'/2 \\
\sigma\tilde\psi + \tilde\psi/2 + \xi\tilde\psi'/2 + \bar\psi\tilde\psi' + \bar\psi'\tilde\psi &= \e^{-\bar\omega}\tilde\omega' - \e^{-\bar\omega}\bar\omega'\tilde\omega + 3\tilde\psi'' + 6\bar\psi'\tilde\omega' + 6\bar\omega'\tilde\psi'.
\end{align}
\end{subequations}
We recover the base state and perturbation to the scaled radius $\phi$ from $\bar \phi = \e^{\bar\omega}$ and $\tilde \phi = \bar\phi\tilde\omega$.  We now consider the computation of the base state and perturbation problems in more detail.

\subsection{Base state}
\label{sec:baseState}

First we examine the base state, which has previously been shown to have discretely selected (and presumably a countably infinite number of) solutions, indexed (say) by decreasing size of $\phi_0$ \cite{Eggers1993, Eggers1995, Brenner1996}.  Rearranging for the derivative terms explicitly:
\begin{subequations}
\label{eqsys:base}
\begin{align}
\bar\omega' &= \frac{1-\bar\psi'/2}{\bar\psi + \xi/2} \\
\bar\psi'' &= \frac{1}{3}\left[\frac{\bar\psi}{2} + \frac{\xi\bar\psi'}{2} + \bar\psi\bar\psi' - \e^{-\bar\omega}\bar\omega' - 6\psi'\bar\omega' \right]
\end{align}
\end{subequations}

As a third order problem, we require three conditions.  Two come from the far field $\xi \to \pm\infty$; per the stationarity condition (that is, that the unscaled velocity remains finite away from the singularity) we must have
\begin{equation}
\label{eq:baseff}
\bar\psi + \xi \bar\psi' \to 0 \quad \Rightarrow \quad \bar\psi \sim C_\pm \xi^{-1}, \quad \xi \to \pm\infty
\end{equation}
where $C_\pm$ are constants that must be determined as part of the solution.  The third comes from the requirement that the solution be analytic near the singular point $\xi_0$ where $\psi = -\xi_0/2$.  This requirement is sensitive as it is a relatively high order noninteger power in the series expansion around $\xi_0$ that can lead to nonanalyticity.  Performing the following expansions
\begin{equation}
\label{eq:baseSeries}
\bar\omega = \sum_{n=0}^\infty c_n(\xi-\xi_0)^n, \qquad \bar\psi = \sum_{n=0}^\infty d_n(\xi-\xi_0)^n,
\end{equation}
we have two free parameters, $\phi_0 = \phi(\xi_0)$ and $\xi_0$, with
\begin{subequations}
\label{eqsys:baseSeriesCoeffs}
\begin{align}
d_0 &= -\xi_0/2, \quad d_1 = 2, \quad c_0 = \log{\phi_0} \\
c_1 &= \frac{\phi_0\xi_0}{12\phi_0-4} \\
d_2 &= -\frac{5c_1}{2} \\
c_2 &= -\frac{c_1^2+6\phi_0}{36\phi_0-2} \\
d_3 &= -\frac{2c_1d_2+10c_2}{3}
\end{align}
\end{subequations}
and so on.  A solution exists when $\phi_0$ and $\xi_0$ are such that both far field conditions are satisfied.

The  boundary value problem is solved in the numerical continuation package AUTO07p \cite{Doedel2007}.  We treat (\ref{eqsys:base}) as two separate boundary value problems from $\xi_0+\delta\xi$ to $L_+\gg 1$, and from $\xi_0-\delta\xi_-$ to $-L_-$ with $L_- \gg 1$.  An initial approximation is made by starting with the values of $\phi_0$ and $\xi_0$ close to those given in \cite{Brenner1996}, and shooting an $O(1)$  distance in each direction.  The power series is used as boundary conditions near the singular point.  After shooting the far field conditions (\ref{eq:baseff}) are not satisfied, a fact that can be quantified using error terms $\epsilon\pm$:
\begin{equation}
\bar\psi_\pm + \xi\bar\psi'_\pm = \epsilon_\pm, \qquad \xi = \pm L_\pm.
\end{equation}
Using numerical continuation we decrease each error $\epsilon_\pm$ to zero, allowing $\xi_0$ and $\phi_0$ to vary, and then increase $L_\pm$ until the finite domain size no longer has a significant effect on the accuracy.

It is important to note that singularities occur when $\phi_0$ is such that the denominator in one of the power series coefficients (\ref{eqsys:baseSeriesCoeffs}) is zero.  At these values, no power series solution exists near $\xi_0$ (instead, the series will presumably have power-log terms). \citet{Brenner1996} observed that these singular values of $\phi_0$ occur at
\begin{equation}
\label{eq:singularphi0s}
\phi_0^* = \frac{1}{15n-12} = \frac{1}{3}, \frac{1}{18}, \frac{1}{33}, \ldots.
\end{equation}
The value of $\phi_0$ corresponding to the first similarity solution, $\phi_0^{(1)} = 0.030432\ldots$, is very slightly larger than the singular value $\phi_0^* = 1/33$, while the value of $\phi_0$ for the second similarity solution $\phi_0^{(2)} = 0.0107854\ldots$ is slightly larger than the singular value $\phi_0^* = 1/93$.  \citet{Brenner1996} argue that the selected values of $\phi_0$ lie close to the singular values \eqref{eq:singularphi0s}.  For this reason it is particularly important to have an appropriate initial approximation for $\xi_0$ and $\phi_0$ for each similarity solution.
In addition, the first nonanalytic term in the series solution of (\ref{eqsys:base}) near $\xi_0$ can be determined to be of order $\xi^{(\phi_0^{-1}+12)/15}$.  The selection of the analytic solution is equivalent to requiring that the coefficient on this term must be zero.  Given (\ref{eq:singularphi0s}), this exponent is marginally less than 3 for the first similarity solution, and marginally less than 7 for the second one.  Terms up to these orders are then needed in the power series to resolve the selection of the first solution. Higher branches of solutions require even further terms.  We go up to 14 terms in this power series expansion (and one term less for the perturbation terms), with the expressions for the coefficients (\ref{eqsys:baseSeriesCoeffs}) found automatically using the symbolic Python library SymPy\,\citep{sympy2017}.

\begin{figure}
\centering
\includegraphics[]{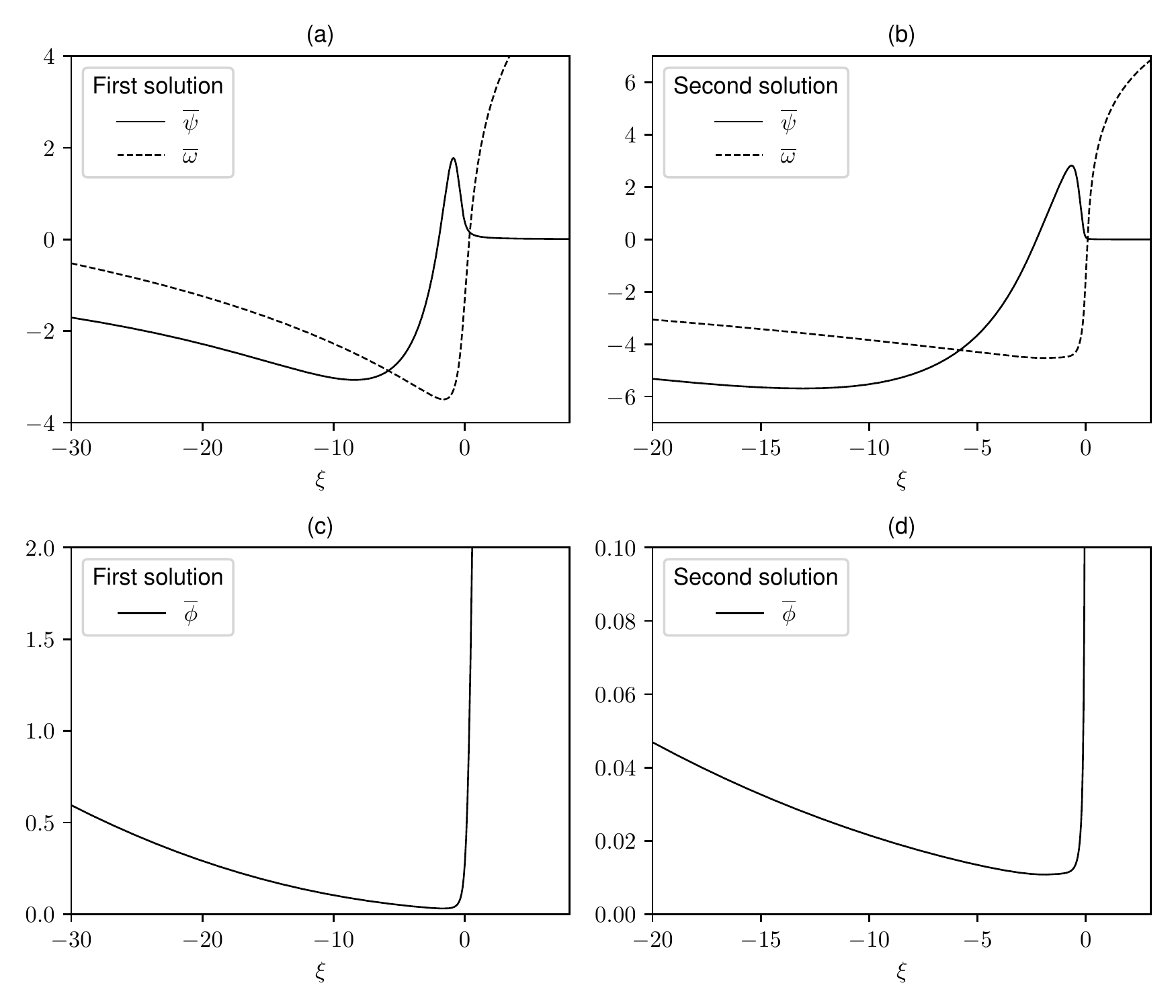}

\caption{(a, c) The first (stable) similarity profile, and (b, d) the second (unstable) similarity profile, solutions to (\ref{eqsys:base}) found by the numerical continuation procedure described in Section \ref{sec:baseState}.  In (c) and (d), the scaled radius $\phi$ is given by $\e^\omega$.}
\label{fig:baseSoln}
\end{figure}

In Fig.~\ref{fig:baseSoln} we plot the first two similarity solutions thus computed (that is, the two solutions to \eqref{eqsys:base} with the largest values of $\phi_0$), which we refer to as the first solution (the solution found by \citet{Eggers1993}, which we expect to be linearly stable), and the second solution.

\subsection{Computation of eigenvalues}

We also find the eigenvalues $\sigma$ using numerical continuation.  Rearranging (\ref{eqsys:corrn0}) for derivatives:
\begin{subequations}
\label{eqsys:corrn}
\begin{align}
\tilde\omega' &= -\frac{\tilde\psi'/2 + \bar\omega'\tilde\psi + \sigma\tilde\omega}{\bar\psi + \xi/2} \\
\tilde\psi'' &= \frac{1}{3}\left[\sigma\tilde\psi + \tilde\psi/2 + \xi\tilde\psi'/2 + \bar\psi\tilde\psi' + \bar\psi'\tilde\psi - \e^{-\bar\omega}(\tilde\omega' - \bar\omega'\tilde\omega) - 6(\bar\psi'\tilde\omega' + \bar\omega'\tilde\psi') \right]
\end{align}
\end{subequations}
Again we require three conditions.  The far field conditions are
\begin{equation}
\label{eq:corrnff}
(2\sigma + 1)\tilde\psi + \xi\tilde\psi' \to 0 \ \Rightarrow \ \tilde\psi \sim \tilde C_\pm\xi^{-(2\sigma+1)}, \qquad \xi \to \pm \infty
\end{equation}
for some constants $\tilde C_\pm$, and again we require analyticity at $\xi_0$.  Expanding the perturbation variables as power series:
\begin{equation}
\label{eq:corrnSeries}
\tilde\omega = \sum_{n=0}^\infty \tilde c_n(\xi-\xi_0)^n, \qquad \tilde\psi = \sum_{n=0}^\infty \tilde d_n(\xi-\xi_0)^n,
\end{equation}
we find the $\tilde d_0$ and $\tilde c_0$ are arbitrary, while the next terms in the power series are found sequentially:
\begin{subequations}
\label{eqsys:corrnSeriesCoeffs}
\begin{align}
\tilde d_1 &= -2\tilde c_0\sigma - 2\tilde d_0c_1 \\
\tilde c_1 &= -\frac{2\tilde d_0\phi_0\sigma+\tilde d_0\phi_0(24c_2+5)+2\tilde c_0c_1}{12\phi_0\sigma+6\phi_0-2} \\
\tilde d_2 &= -\frac{2\tilde c_1\sigma+4\tilde d_0c_2+2c_1\tilde d_1+5\tilde c_1}{2} \\
\tilde c_2 &= -\frac{\phi_0\tilde d_1\sigma+36\tilde d_0\phi_0c_3+2\tilde d_0\phi_0d_2+(12\phi_0\tilde d_1+2\tilde c_0)c_2+5\phi_0\tilde d_1+2c_1\tilde c_1-\tilde c_0c_1^2}{12\phi_0\sigma+36\phi_0-2} \\
\tilde d_3 &= -\frac{2\tilde c_2\sigma+6\tilde d_0c_3+2c_1\tilde d_2+2\tilde c_1d_2+10\tilde c_2+4\tilde d_1c_2}{3}
\end{align}
\end{subequations}
and so on.  Expressions for these coefficients up to 14th order are again found with SymPy\,\cite{sympy2017}.  

When considering the equations for the perturbations, the base state is fixed, thus $\phi_0$ is a given value, and the effect of varying $\sigma$ must be considered.  In a similar manner to the leading order case, the perturbation problem has no analytic solution when a denominator in the power series coefficients for $\tilde\omega$, that is, (\ref{eqsys:corrnSeriesCoeffs}b,d), etc., is zero.  These occur at specific values $\sigma = \sigma^*$:
\begin{equation}
\label{eq:singularSigmas}
\sigma^* = \frac{1}{6}\left(\phi_0^{-1} - (15n-12)\right), \qquad n = 1, 2, 3, \ldots.\end{equation}
Based on the value $\phi_0^{(1)} = 0.030432$, the singular values of $\sigma$ for the first solution occur at
\begin{subequations}
\label{eqsys:singulareigs}
\begin{equation}
\sigma^{*(1)} = 4.976630,   2.476630,  -0.023370,  -2.523370, \ldots,
\end{equation}
while on the second solution, $\phi_0^{(2)} = 0.107854$ and thus the singular values of $\sigma$ are
\begin{equation}
\sigma^{*(2)} = 14.953041, 12.453041,  9.953041,  7.453041,  4.953041,
        2.453041, -0.046959,  \ldots.
\end{equation}
\end{subequations}

The system (\ref{eqsys:corrn}) is coded into AUTO alongside the system for computation of the base state, with (\ref{eq:corrnff}) and (\ref{eq:corrnSeries}) as boundary conditions, but allowing $\tilde c_0 = \tilde c_{0\pm}$ and $\tilde d_0 = \tilde d_{0\pm}$ to take different values on either side of $\xi_0$ in the power series.  As the perturbation problem is linear, the scale invariance must be removed by fixing either of $\tilde c_0$ and $\tilde d_0$.  If $\tilde d_0$ is fixed, then for given $\sigma$, $\tilde c_{0\pm}$ is determined by satisfying each far field condition. 
The artificial degree of freedom in allowing $\tilde c_{0}$ to take different values on each side of the singular point allows us to vary $\sigma$ via continuation, and detect discrete eigenvalues by the criterion $\tilde c_+ - \tilde c_- = \delta\tilde c = 0$ (that is, when $\delta\tilde c = 0$, the two solutions on either side of $\xi_0$ have been connected in a continuous and smooth fashion).  Conversely, if $\tilde c_0$ is fixed, we detect eigenvalues by allowing $\tilde d_{\pm}$ to vary, and then the values of $\sigma$ such that $\delta\tilde d= \tilde d_{0+} - \tilde d_{0-} = 0$ correspond to eigenvalues.

\subsubsection{Eigenvalues on the real line}

We start by computing real eigenvalues $\sigma$ for both first and second solutions.  Due to invariance in space and time of the original problem, respectively, we expect trivial unstable eigensolutions:
\begin{subequations}
\label{eqsys:trivialEigs}
\begin{align}
\sigma_X &= \frac{1}{2}, \qquad \tilde\omega_X = \bar\omega', \qquad \tilde\psi_X = \bar\psi', \label{eq:trivialEigX} \\
\sigma_T &= 1, \qquad \tilde \omega_T = \xi\bar\omega' - 2, \qquad \tilde\psi_T = \bar\psi + \xi\bar\psi', \label{eq:trivialEigT}
\end{align}
up to a scalar constant of multiplication in the eigenfunctions \cite[e.g.][]{Giga1985,Witelski1999}.  In addition, there is a stable eigenvalue due to Galilean invariance under constant-speed translation, i.e. $h(z,t) \mapsto h(z-\epsilon(t-t_0),t)$, $v(z,t) \mapsto v(z-\epsilon(t-t_0), t) + \epsilon$.  Under this translation we find the eigensolution
\begin{equation}
\label{eq:trivialEigV}
\sigma_V = -\frac{1}{2}, \qquad \tilde\omega_V = \bar\omega', \qquad \tilde\psi_V = 1+\bar\psi'.
\end{equation}
\end{subequations}
While these eigenvalues do not play a role in the stability of similarity solutions (these instabilities are removed by correctly choosing the time and location of the break-up point), and (as we will see) real eigenvalues do not play a role in the stability of the similarity solutions we consider here, the accurate computation of the trivial eigenvalues provides a useful check on the accuracy of our method.

The real eigenvalues for both the first and second similarity solutions are found as follows.  Starting with an initial guess of zero for the eigenfunctions and a given value of $\sigma$, the parameter $\tilde d_0$ is brought from zero to unity, while allowing $\tilde c_{0+}$ to vary, as well as the difference $\delta\tilde c$ between the values on each side of $\xi_0$.  
By varying $\sigma$ and computing the corresponding value for $\delta\tilde c_0$, we construct a selection curve; when $\delta\tilde c_0 = 0$, we have a candidate eigenvalue.  Note that by fixing $\tilde d_0$, we remove the arbitrary constant in the linear problem, at the cost of introducing singularities when a solution requires $\tilde d_0 = 0$.  We build a complete curve by choosing values of $\sigma$ on either side of each singularity to continue from.  If we instead fix the value of $\tilde c_0$ and let $\tilde d_{0\pm}$ be free, the singularities occur in different places, while the same zeros (corresponding to the eigenvalues) are found.

\begin{figure}
\centering
\includegraphics[]{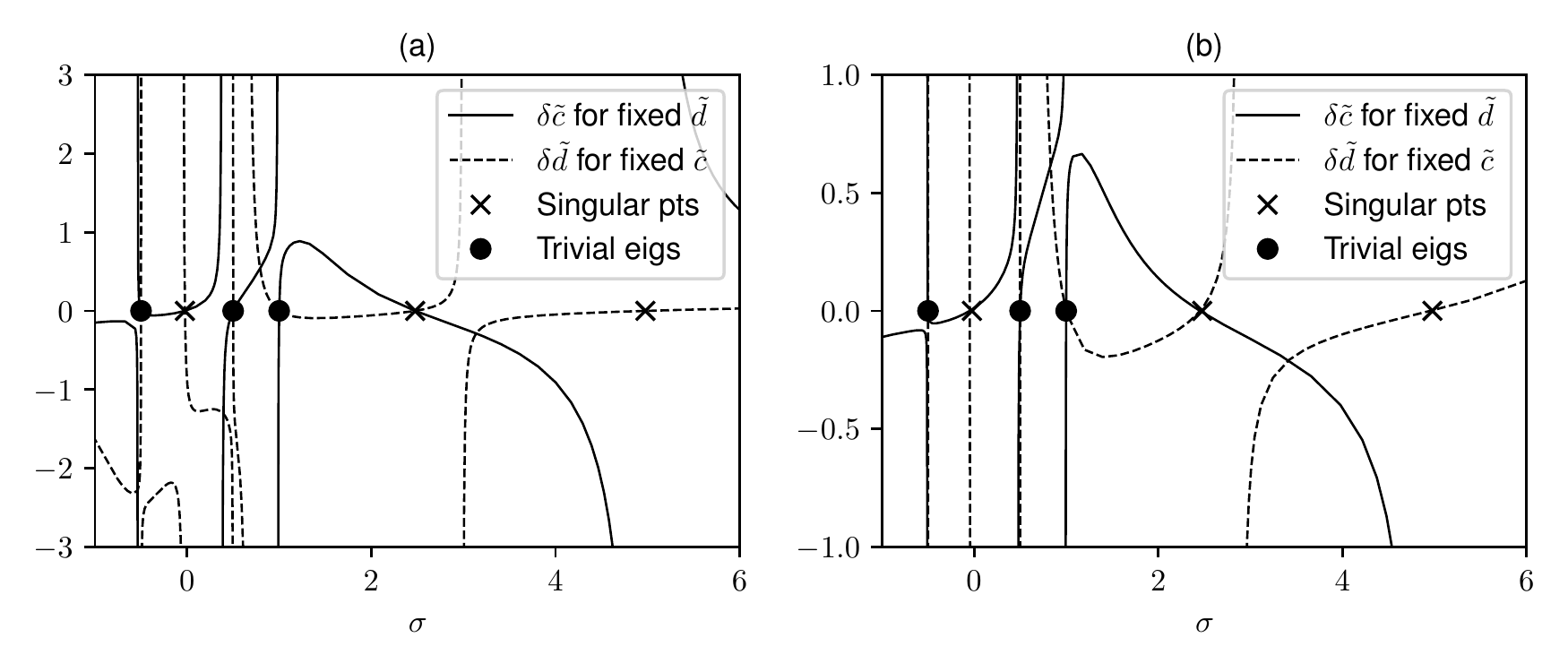}

\caption{Selection of eigenvalues $\sigma$ on the real line: (a) the first solution, and (b) the second solution.  We see the three trivial eigenvalues at $\sigma_T = 1$, $\sigma_X=1/2$, and $\sigma_V=-1/2$, as well as the singular values $\sigma^*$ (\ref{eqsys:singulareigs}), where a coefficient in the power series expansion for the perturbation (\ref{eqsys:corrnSeriesCoeffs}) becomes singular, and therefore do not correspond to an analytic solution.}
\label{fig:select}
\end{figure}

The resultant selection curve is shown in Fig.~\ref{fig:select} for both the first and second solutions, and showing the results both of fixed $\tilde c$ and of fixed $\tilde d$.  As expected, the three trivial eigenvalues (\ref{eqsys:trivialEigs}) at $\sigma = 1/2, 1$ and $-1/2$ are observed.  In Fig.~\ref{fig:select} we plot the results between $\sigma=-1$ and $\sigma=6$, showing no other real eigenvalues exist in this region; we have continued for larger values up to $\sigma = 100$ and not found any other eigenvalues, and have no reason to suspect they may exist, given the absence of the large exponential growth this would predict in the simulations (see section \ref{sec:numerics}).  Negative real eigenvalues may exist for $\sigma < -1$, but these are not important compared to the complex eigenvalues we describe in the next section.

We also note the existence of apparent eigenvalues at the values $\sigma^*$ (\ref{eq:singularSigmas}), at which point the power series for the perturbation solution at $\xi_0$ does not exist.  These points must be discounted as the eigenfunction is not analytic at $\xi_0$ for these values (the continuation method has managed to pass through these points numerically).  More generally, any method of analyzing the linear stability of this problem is likely to come up with these spurious values, demonstrating that care must be taken in interpreting the computation of such points with regard to the actual linear stability.  As will become apparent when comparing with numerical solutions, these singular values $\sigma^*$ do not correspond to unstable modes.

\subsubsection{Complex eigenvalues}
All of the eigenvalues found thus far are trivial, and will not determine the stability of a similarity solution. Instead, we must look for eigenvalues in the complex $\sigma$-plane.  We now search for complex eigenvalues by assuming $\tilde \omega$, $\tilde \psi$ (and thus $\tilde c_0 = \tilde c_{0R} + \ii \tilde c_{0I}$ and $\tilde d_0 = \tilde d_{0R} + \tilde d_{0I}$) and $\sigma = \sigma_R + \mathrm i\sigma_I$ to be complex, and solving both real and imaginary parts of the equations (\ref{eqsys:corrn}), as well as conditions (\ref{eq:corrnff}) and (\ref{eq:corrnSeries}).  

To locate complex eigenvalues, we fix $\tilde d_{0R}$ and $\tilde d_{0I}=0$, and then must find $\sigma_R$ and $\sigma_I$ such that both $\tilde c_{0R}$ and $\tilde c_{0I}$ vanish.  To accomplish this, we start with a range of real values of $\sigma$, and $\tilde d_0$ is again increased to a fixed value to fix the arbitrary constant in the eigenfunction.  Then, $\sigma_I$ is increased until the real part of the error $\delta \tilde c_{0R} = 0$, at which point the curves defined by $\delta c_{0R} = 0$ are traced out to find points where the imaginary part of the error $\delta \tilde c_{0I} = 0$ as well.  This point then corresponds to a complex eigenvalue.  By starting with a large number of values of $\sigma_R$, we can be confident of finding all such curves and eigenvalues in some region near the real line.

\begin{figure}
\centering
\includegraphics[]{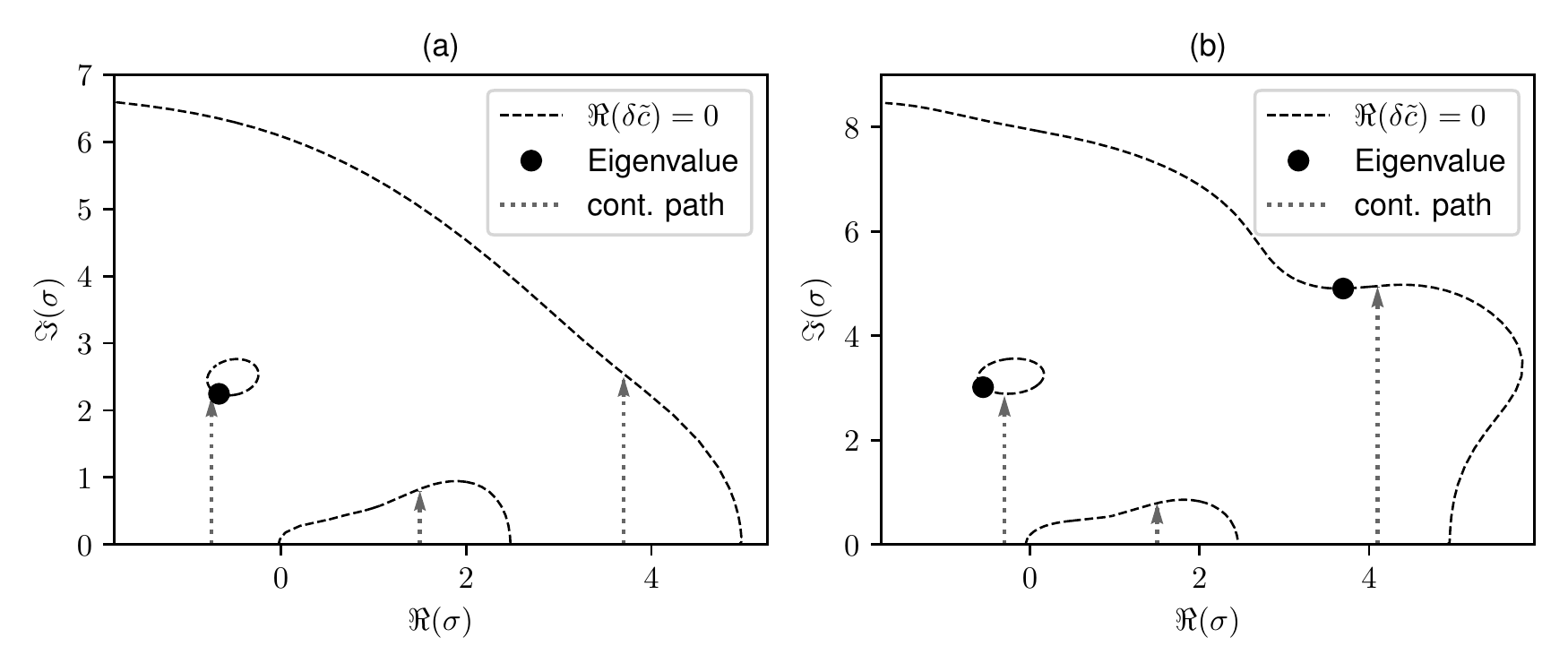}

\caption{Curves defined by the real part of the error parameter $\Re(\delta\tilde c) = \delta \tilde{c}_R=0$ and eigenvalues (where both real and imaginary parts of $\delta\tilde c$ vanish) for the (a) first solution at $\sigma = -0.670 + 2.246\ii$ (b) second solution at $\sigma = -0.552 + 3.015\ii, $ and $\sigma = 3.695 + 4.905\ii$.  In both cases, the eigenvalues are found by continuing $\sigma$ from real values in the positive imaginary direction, until the real part of the error $\delta\tilde c_{0R}$ is zero, then continuing into the complex plane (holding $\delta\tilde c_{0R}$ at zero) to find the curves defined by this equation.  At points on these curves where the error $\delta\tilde c_{0I}$ is also zero, the corresponding value of $\sigma$ is an eigenvalue.}
\label{fig:eig}
\end{figure}

\begin{figure}
\centering
\includegraphics[]{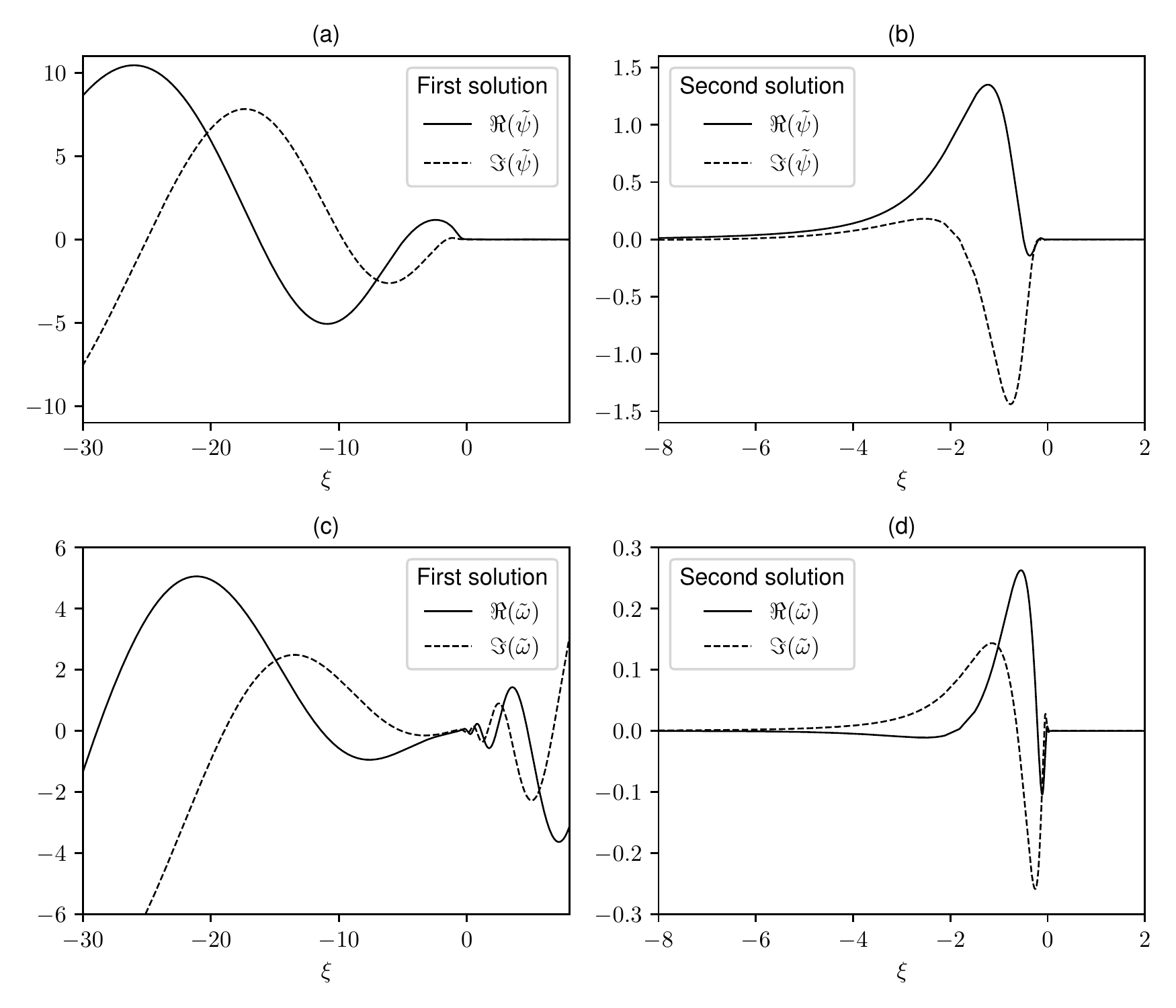}

\caption{The complex eigenfunction for each of (a, c) the stable first similarity solution at the complex eigenvalue $\sigma = -0.670 + 2.246\ii$, and (b, d) the unstable second similarity solution at $\sigma = 3.695 + 4.905\ii$.  (a) and (c) plot the perturbations $\tilde\psi$ to the self-similar velocity profiles $\psi$, while (b) and (d) are the perturbations $\tilde w$ to the self-similar profile of the log-radius $w$.  The unstable eigenfunctions on the second solution are localized in the region near zero which includes both the singular point and the region in which the base solution (depicted in Fig.~\ref{fig:baseSoln}b,d) have large gradients.}
\label{fig:complexEigSoln}
\end{figure}

These curves and eigenvalues are plotted in Fig.~\ref{fig:eig}.  %
For the first similarity solution, the eigenvalue with largest real part thus found occurs at $\sigma \approx -0.670 + 2.246\ii$ (of course, the complex conjugate is also an eigenvalue).  The location of this eigenvalue is plotted in Fig.~\ref{fig:eig} along with the continuation path taken to find it, while the eigenfunctions $\tilde\omega, \tilde\psi$ themselves are plotted in Fig.~\ref{fig:complexEigSoln}.  Since this eigenvalue has negative real part, the first similarity solution is linearly stable.  This conforms with numerous previous numerical observations\cite{Eggers1993,Eggers1995,Brenner1996}); in Section \ref{sec:numerics} we show the complex value of $\sigma$ accurately predicts both the frequency and decay rate of the original system (\ref{eqsys:pdes}) as the stable similarity solution is approached.  In addition, since the real part of this eigenvalue is greater than $-1$, the effect of this eigenvalue is dominant over the effect of higher order contributions to the surface tension (proportional to $(t_0-t) = \e^{-\tau}$) that were neglected in deriving the system (\ref{eq:scaledpde2}).

For the second similarity solution, complex eigenvalues are found in the same way. The two nontrivial eigenvalues with largest real part are thus found, one at $\sigma = -0.552 + 3.015\ii$, and one at $\sigma = 3.695 + 4.905\ii$.  The presence of an eigenvalue with positive real part indicates the second similarity solution is linearly unstable.  The eigenfunctions are again plotted in Fig.~\ref{fig:complexEigSoln}.  
These results comprise the first direct evidence that it is purely linear stability considerations that are the reason that only the first similarity solution is seen in reality and numerical simulations, and not finite-size effects as conjectured by \cite{Brenner1996}.  An important property of the unstable eigenfunction (Fig.~\ref{fig:complexEigSoln}b,d) is that it is localized near the region where the gradient of the base velocity is large, and the gradients of the eigenfunctions are themselves large.  Since the WKB analysis in \cite{Brenner1996} presupposes gently sloping $\phi, \psi$, with a relatively large wavenumber, it is not surprising that their analysis does not detect this instability. A discussion on the convergence of the eigenvalues with respect to the numerical parameters, particularly $\delta\xi$, is included in Appendix~\ref{app:convergence}.

\section{Comparison with numerical simulation}
\label{sec:numerics}

\subsection{Numerical scheme}
We now show that the numerical solution demonstrates the validity of the linear stability results.   The numerical solutions of full nonlinear lubrication equations (\ref{eqsys:pdes}) are computed using a second-order finite difference scheme in both time and space. The time-integration method is
an adaptive Runge--Kutta scheme, where the time step is adjusted so that second-order temporal derivative in the Taylor expansion $<10^{-3}$.
The spatial grids are highly non-uniform, with corresponding finite-difference weights calculated by Fornberg algorithm \cite{fornberg1998classroom}, which enables high resolution of the region near the rupture point. When necessary, additional mesh points are added to the grid at points where the minimum height is smaller than a specified percentage of the local grid size, with the solution interpolated onto the new grid using cubic splines.

\subsection{Decay to stable similarity solution}

\begin{figure}
\centering
\includegraphics[]{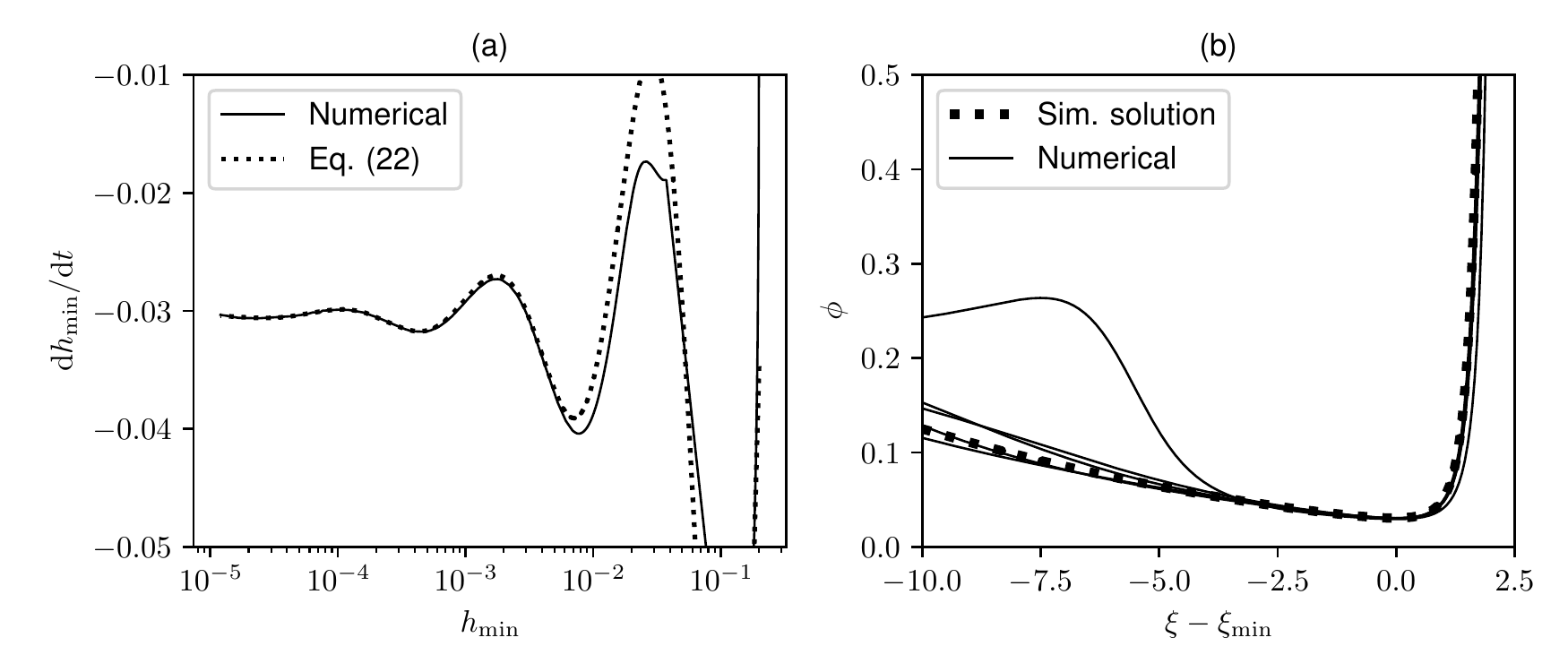}

\caption{Convergence of a generic initial condition to the stable similarity solution: (a) By plotting $h_\mathrm{min}$ against its time derivative and comparing to (\ref{eq:dhmindt}) the decay rate and frequency predicted by the dominant nontrivial eigenvalue $\sigma \approx -0.670 + 2.246\ii$ can be clearly observed (the values $\mathcal A_1 = 0.25$ and $\mathcal A_2 = 2$ are fitted); (b) The profiles asymptote to the stable similarity solution as time tends to the pinch-off time (the profiles are shifted by the point $\xi_\mathrm{min}$ where $\phi$ is at its minimum, so the shifted profiles all have their minimum at zero).}
\label{fig:stableComparison}
\end{figure}

In Fig.~\ref{fig:stableComparison} we plot the result for a generic initial condition, showing (Fig.  \ref{fig:stableComparison}a) the oscillatory decay to the stable similarity solution, and (Fig. \ref{fig:stableComparison}b) the asymptotic agreement with the similarity profile.  By plotting $\log (dh_{\mathrm{min}}/dt)$ as a function of $\log h_\mathrm{min}$ in Fig.~\ref{fig:stableComparison}a, we observe the convergence to the similarity solution without having to estimate the break-up time and location explicitly, as has been exploited in previous studies \cite{Li2016}.
In terms of the similarity variables, $h_\mathrm{min}$ occurs at a point $\xi^*(\tau)$, where 
\begin{equation}
h_z(z,t) = (t_0-t)^{1/2}\phi_\xi(\xi^*(\tau),\tau) = 0.
\end{equation}
As the singularity is approached for a stable solution, $\tau\rightarrow\infty$ and $\xi^*(\tau)$ converges toward the minimum $\bar{\xi}^*$ of the similarity profile $\bar{\phi}(\xi)$. Note that $\bar\xi^*$ is close to, but not equal to the singular point $\xi_0$.  Assume in accordance with linear stability that $\phi \sim \bar\phi(\xi) + c\e^{-\sigma\tau}\tilde\phi(\xi)$ (plus complex conjugate), where $c \in \mathbb C$. We then have
\begin{equation}
\xi^*(\tau) \sim \bar\xi^* + c\e^{-\sigma\tau}\tilde\xi^*, \qquad \tilde \xi^* = -\frac{\tilde\phi'(\bar\xi^*)}{\bar\phi''(\bar\xi^*)},
\end{equation}
and
\begin{equation}
\label{eq:dhmindt}
\frac{\mathrm dh_\mathrm{min}}{\mathrm dt} \sim -\bar\phi(\bar\xi^*) +  \mathcal A h_\mathrm{min}^{-\sigma}. + \mathrm{c.c.} = -\phi_\mathrm{min} + \mathcal A_1h_\mathrm{min}^{-\sigma_R}\cos\left(\sigma_I\log(h_\mathrm{min}) + \mathcal A_2\right), \quad \mathcal A_1, \ \mathcal A_2 \in \mathbb R
\end{equation}
where 
\begin{equation}
\label{eq:const_in_hmin_relationship}
\mathcal A = \mathcal A_1\e^{-\ii\mathcal A_2} =  c(\sigma - 1)\bar\phi(\bar\xi^*)^\sigma\tilde\phi(\bar\xi^*).
\end{equation}
Note that since $\mathcal A = 0$ when $\sigma = 1$ the instability due to time invariance (\ref{eq:trivialEigT}) is suppressed, while the instability due to translation invariance (\ref{eq:trivialEigX}) and the (stable) perturbation due to velocity invariance (\ref{eq:trivialEigV}) are suppressed since the corresponding eigenfunctions have $\tilde\phi = \bar\phi\tilde \omega = \bar\phi' = 0$ at $\xi = \bar \xi^*$.  Thus the dominant behavior is due to the nontrivial eigenvalue with largest real part.

The amplitude $\mathcal A_1$ and phase $\mathcal A_2$ in (\ref{eq:dhmindt}) are not predicted from linear stability, but the decay rate and frequency are given by the real and imaginary parts of the eigenvalue $\sigma$.  In Fig.~\ref{fig:stableComparison} we plot the predicted curve (\ref{eq:dhmindt}), with $\mathcal A_1, \mathcal A_2$ suitably fitted.  The agreement between the linear stability result and the numerical computation is excellent, and it is indeed the complex eigenvalue computed in Section \ref{sec:similarity} that dominates the dynamics as the similarity solution is approached.

A challenge of comparing numerical solutions to similarity solutions representing finite-time singularities (such as the pinch-off in the system (\ref{eqsys:pdes})) is that any small but finite (and hence beyond linear stability analysis) source of error, such as the spatial discretization, becomes increasingly important as the singularity time is approached.  In our numerical scheme we found that the effect of the error is to destabilize the stable similarity solution; however, we observe that the time that this destabilization becomes appreciable can be made arbitrarily close to the pinch-off time by decreasing the mesh size, thus the destabilization is a property of the numerical scheme and not of the PDE system (\ref{eqsys:pdes}) itself.  This is however reminiscent of the observation that physically, any finite-size noise, for example due to thermal fluctuations (\cite{Eggers2002, Zhao2019}) will destabilize the similarity solution at very small scales \cite{Brenner1994}.

\subsection{Linear instability of second solution}

\begin{figure}
\centering
\includegraphics[]{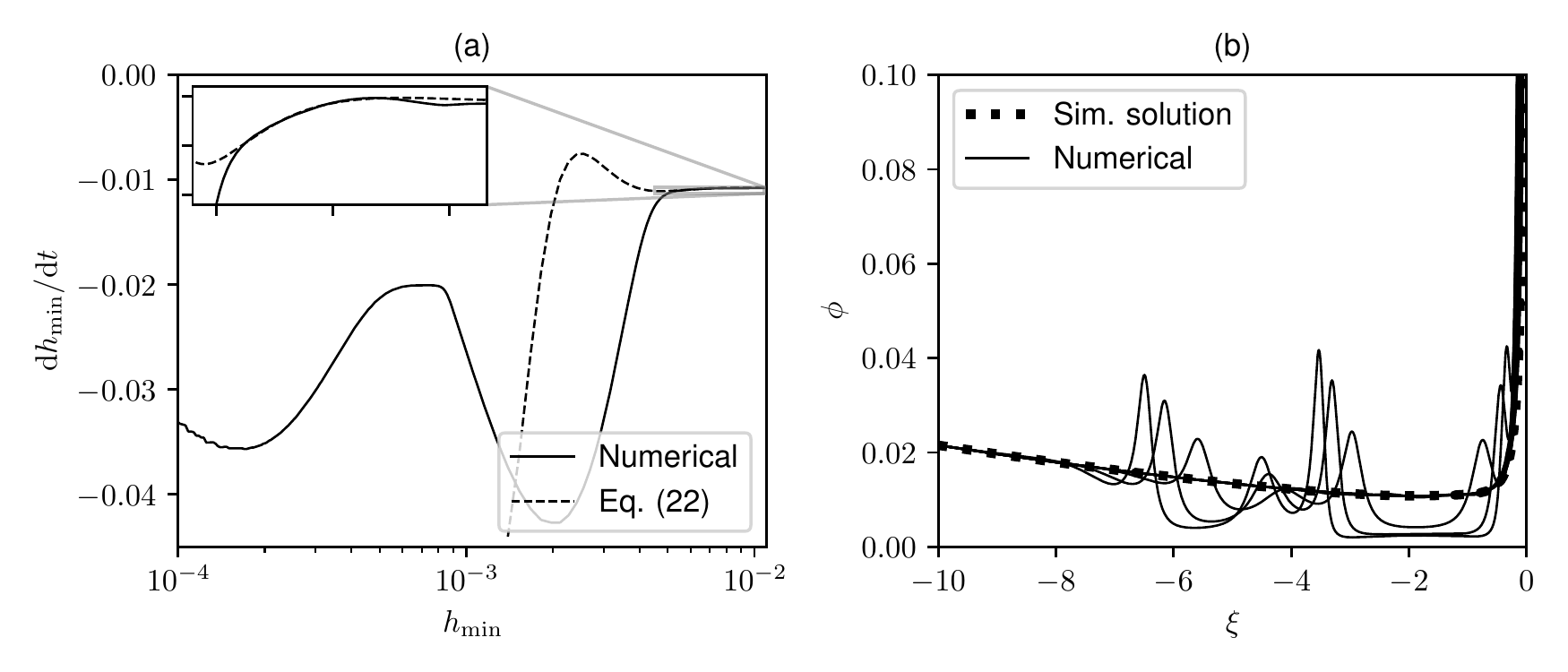}

\caption{Departure of the numerical solution to \eqref{eqsys:pdes} from an initial condition equal to the second similarity solution.  (a) the plot of $h_\mathrm{min}$ against its derivative shows the solution leaving the neighborhood of the unstable solution and settling to the stable solution (where $\mathrm dh_\mathrm{min}/\mathrm dt \approx -0.0304$).  The inset shows a short-lived, but reasonable, match to the linear stability approximation \eqref{eq:dhmindt}.  (b) The profiles show the oscillations grow first near the minimum in $\phi$.}
\label{fig:unstablecomparison}
\end{figure}

In Fig.~\ref{fig:unstablecomparison} we plot the result of the numerical simulation with initial condition given by the unstable similarity solution plotted in Fig.~\ref{fig:baseSoln}(b,d).  By equating $h(x,0)$ with $\phi(\xi)$ and $v(x,0)$ with $\psi(\xi)$, the time to rupture would (if the solution remained at the similarity solution) be $t_0 = 1$.

A quantitative comparison with the linear stability prediction is challenging, for a number of reasons; the full numerical solution includes the nonzero contribution from the higher order surface tension from \eqref{eqsys:pdes}, which does not appear in the similarity solution, which will have an effect at early time, and since the minimum of $\phi$ is close to zero while the real part of the eigenvalue is large, we expect the solution to rapidly leave the linear regime.   In Fig.~\ref{fig:unstablecomparison}a we plot $h_\mathrm{min}$ against its time derivative, with a fit of the predicted behavior according to \eqref{eq:dhmindt}.  A meaningful fit can only be made over a relatively short range (the inset of Fig.~\ref{fig:unstablecomparison}(a)), which, although inconclusive, is expected for the above reasons.

The solution (scaled according to the similarity variables) are plotted in Fig.~\ref{fig:unstablecomparison}b.  The profiles remain close to the unstable base state for some time, before rapidly thinning near the minimum in $\phi$.

\section{Discussion}
\label{sec:discussion}

In this article we have shown by explicit computation of the linear stability problem in the appropriate self-similar coordinate system (\ref{eqsys:scaledpdes}) that the first solution (ordered by largest minimum thickness) is linearly stable, while the second solution is linearly unstable.  The linear stability computations accurately determine the trivial eigenvalues that are known to exist, and match with the behavior of numerical solutions, providing further evidence of their accuracy.  Although we have only computed the linear stability properties of the first two similarity solutions, there is no reason to believe that the more extreme solutions found by \cite{Brenner1996} will be linearly stable, as they are also not seen in simulations. 

The existence of an oscillatory decay to the stable similarity solution was first noted in \citet{Li2016}, who conjectured the existence of the dominant complex eigenvalue we have found in this study.  As noted in \citet{Li2016}, this oscillatory decay has not been emphasized in most experimental and numerical studies, even where it is likely to be present, as it is not obvious when plotting the minimum radius against time to pinch-off, rather than the time derivative of minimum radius against minimum radius, as in Fig.~\ref{fig:stableComparison}a.  In addition, most experiments start in a regime where the inertial or viscous effects are dominant, and it is only in the stages very close to pinch-off that the viscous-inertial regime is reached.

However, our linear stability analysis does suggest the oscillatory decay is the generic behavior as a solution evolves from another profile (for instance, as it would when transiting from the viscous to the viscous-inertial regime), and at least one or two oscillations could have a large enough amplitude to have an important physical effect.  One intriguing possibility is that it is these decaying oscillations that are the cause of small bumps or oscillations that are sometimes observed on the `micro-threads' close to pinch-off (see Fig.~8 of \citet{Kowalewski1996}).  These small bumps cause very small satellite droplets to be created when a larger droplet pinches off~\citep{Castrejon2015, Latikka2020}.  
A reason to believe this possibility is that the small bumps do not exhibit a well-defined critical wavelength, as might be expected from a Rayleigh-Plateau type of instability in non-self-similar coordinates, and the small droplets decrease in size the closer they are to the main pinch-off location.  We note that as the oscillations are decaying in the self-similar coordinate system, we would not expect the droplets created to be strictly geometrically decreasing in size, as might occur if a periodic solution in self-similar coordinates existed (as occurs in the thin film equation of \citet{Dallaston2018}, for example).  However, the presence of oscillations could create further pinch-off locations that lead to a large distribution of final droplet sizes.

It is hoped that this study will help to clarify what parts of the dynamics of near-pinch-off behavior of liquid threads can be explained by the continuum dynamics described by the system (\ref{eqsys:pdes}), and what parts would require extra effects such as the viscosity of the ambient fluid, or thermal or other small scale fluctuations, to explain.  We have also outlined how numerical continuation may be used to numerically compute challenging linear stability problems, particularly those that have singular points in the domain interior.  

\begin{acknowledgments}
CZ and JES acknowledge financial support from EPSRC grants EP/N016602/1, EP/P020887/1, EP/S029966/1, and EP/P031684/1.
\end{acknowledgments}

\begin{appendix}
\section{Convergence of eigenvalues with respect to $\delta\xi$}
\label{app:convergence}
 We found the solutions to both the base state (\ref{eqsys:base}) and perturbation problems (\ref{eqsys:corrn}) to be well converged with a value of the distance from the singular point at which the power series is applied $\delta\xi = 0.2$ and domain size $L^- = 50$, $L^+ = 10$ for the first branch, and $\delta\xi = 0.3$ and $L^- = 10$ and $L^+ = 5$ for the second branch (these have been sometimes increased for the sake of illustration).  For these values, even the least accurately determined of the trivial eigenvalues ($\sigma_V = -1/2$ on the second branch) was correct to three decimal places, with the other trivial eigenvalues found with greater accuracy. 
 
\begin{figure}
\centering
\includegraphics[]{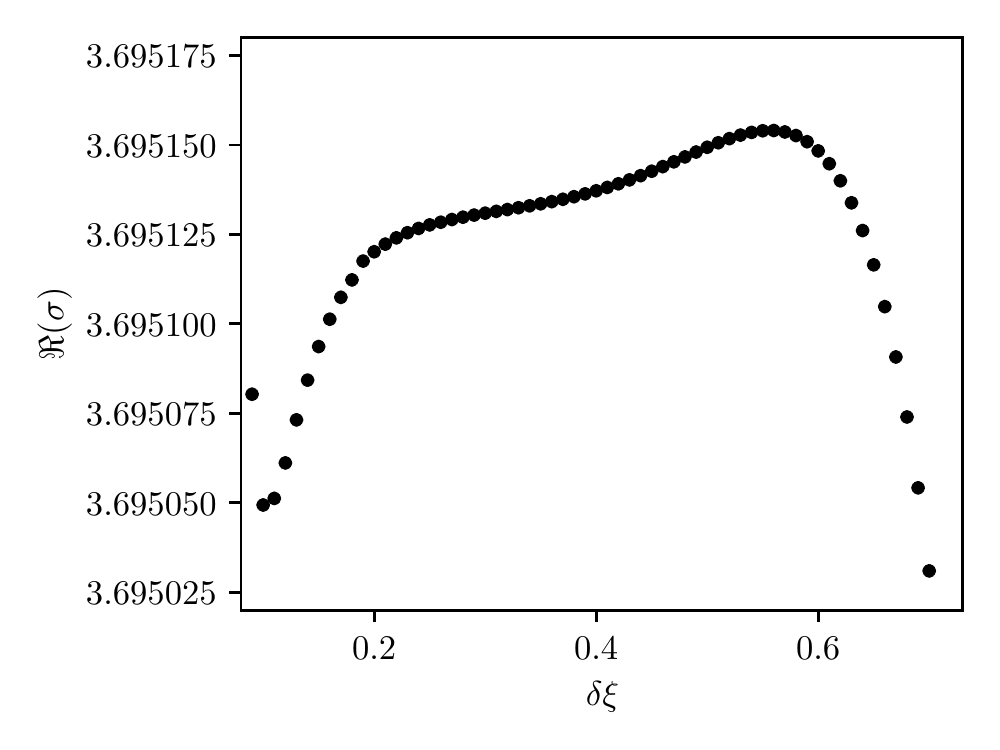}
 
\caption{The convergence of the system at the unstable complex eigenvalue $\sigma = 3.695 + 4.905\ii$. As $\delta\xi$ decreases, the solution is observed to converge, until $\delta\xi$ tends to below $0.1$, where numerical round-off error in the high order power series starts to have an effect.  Nevertheless, the eigenvalue has been accurately determined to at least three decimal places.}
\label{fig:convergenceIndxi}
\end{figure}

To further ensure the accuracy of the nontrivial complex eigenvalues, we have checked the convergence with respect to the numerical parameters, in particular $\delta\xi$: the distance from the singularity at which the power series expansions (\ref{eqsys:baseSeriesCoeffs}) and (\ref{eqsys:corrnSeriesCoeffs}) are applied.  The convergence in this parameter is complicated by the sensitivity of the selection of solutions (i.e., that it is the existence of higher derivatives that determine the existence of a solution), particularly the second solution.  For small $\delta\xi$, accuracy is lost due to numerical round-off error in the calculation of coefficients in (\ref{eqsys:baseSeriesCoeffs}) and (\ref{eqsys:corrnSeriesCoeffs}), as the terms responsible for selection become too small.  In Fig.~\ref{fig:convergenceIndxi} we plot the unstable eigenvalue on the second solution as this parameter is varied.  This plot shows that accuracy is lost if this parameter is too big or small, but as demonstrated, for a range of values between $\delta\xi = 0.2$ to $0.6$, the eigenvalue has converged to three decimal places.

\end{appendix}

\end{document}